# Magnetic field strength influence on the reactive magnetron sputter deposition of $Ta_2O_5$

**R Hollerweger[1,3], D Holec[2], J Paulitsch[1,3], R Rachbauer[4], P Polcik[5] and P H Mayrhofer[1,3]**

[1] Christian Doppler Laboratory for Application Oriented Coating Development at the Department of Physical Metallurgy and Materials Testing, Montanuniversität Leoben, Franz-Josef-Str. 18, A-8700 Leoben, Austria

[2] Department of Physical Metallurgy and Materials Testing, Montanuniversität Leoben, Franz-Josef-Str. 18, A-8700 Leoben, Austria

[3] Institute of Materials Science and Technology, Vienna University of Technology, A-1040 Vienna, Austria

[4] OC Oerlikon Balzers AG, Iramali 18, LI-9469 Balzers, Principality of Liechtenstein

[5] Plansee Composite Materials GmbH, Siebenbürgerstr. 23, D-86983 Lechbruck am See, Germany

robert.hollerweger@tuwien.ac.at

**Abstract.** Reactive magnetron sputtering enables the deposition of various thin films to be used for protective as well as optical and electronic applications. However, progressing target erosion during sputtering results in increased magnetic field strengths at the target surface. Consequently, the glow discharge, the target poisoning, and hence the morphology, crystal structure and stoichiometry of the prepared thin films are influenced. Therefore, these effects were investigated by varying the cathode current $I_m$ between 0.50 and 1.00 A, the magnetic field strength B between 45 and 90 mT, and the $O_2/(Ar+O_2)$ flow rate ratio $\Gamma$ between 0 and 100%. With increasing oxygen flow ratio a sub-stoichiometric $TaO_x$ oxide forms at the metallic Ta target surface which further transfers to a non-conductive tantalum pentoxide $Ta_2O_5$, impeding a stable DC glow discharge. These two transition zones (from Ta to $TaO_x$ and from $TaO_x$ to $Ta_2O_5$) shift to higher oxygen flow rates for increasing target currents. Contrary, increasing the magnetic field strength (e.g., due to sputter erosion) mainly shifts the $TaO_x$ to $Ta_2O_5$ transition to lower oxygen flow rates while marginally influencing the Ta to $TaO_x$ transition. To allow for a stable DC glow discharge (and to suppress the formation of non-conductive $Ta_2O_5$ at the target) even at $\Gamma = 100\%$ either a high target current ($I_m \geq 1$ A) or a low magnetic field strength ($B \leq 60$ mT) is necessary. These conditions are required to prepare stoichiometric and fully crystalline $Ta_2O_5$ films.

## 1. Introduction

Reactive magnetron sputtering is a widely spread technique to deposit thin compound films on a huge variety of substrate materials. However, to reproducibly deposit high quality coatings a consolidated knowledge on process influencing effects like the target poisoning behaviour and its impact on the plasma conditions is necessary [1–6]. Therefore, many investigations are focusing on improving process controlling parameters such as power supply, power density, reactive gas flow, total pressure, pumping speed or magnetic field strength [2,7,8].

Determining a cathode-voltage-hysteresis as a function of the reactive gas flow gives an easy access to investigate the poisoning behaviour of the cathode material. In general two operating modes can be distinguished: The metallic sputtering mode, established at low reactive gas contents, in which the target surface still indicates the metallic phase, and the poisoned sputtering mode at higher reactive gas contents, in which a compound layer is formed at the target surface [9,10]. Such compounds can negatively affect the discharge properties due to changes in sputter yields, secondary electron emission or conductivity [11,12].

Magnetron sputtering was introduced in the 70ies as it enables a stable glow-discharge at lower chamber pressures than known for diode-sputtering [13]. The magnetic field of the magnetron is responsible for trapping secondary electrons, which are essential for ionizing the gaseous species and consequently for the plasma density. Nevertheless, progressing sputter erosion of the target material results in an increase of the magnetic field strength and hence to a changed discharge behaviour of the target material [14,15].

Tantalum pentoxide, $Ta_2O_5$, is a promising candidate for optical, electronic, corrosion-protective and biocompatible applications [16–18]. However, the reactive sputtering process is very sensitive to the reactive gas partial pressure used. Therefore Schiller et al. have studied the influence of the sputtering power, total pressure as well as partial pressure on the sputter rate of Ta and Ti in an $Ar/O_2$ glow discharge. Based on their results, they have derived a model for the formation of a microscopic (conductive) or macroscopic (insulating) oxide-compound layer on the target surface to describe the radial oxide coverage and the discharge area [19,20].

Within this study, we investigate the influence of the magnetic field strength, the sputtering current and the oxygen gas flow on the poisoning behaviour of a tantalum target. Our results emphasize the need of a well-balanced process especially when considering the change in magnetic field strength due to continuous sputter erosion of the target surface and high oxygen flow rates which are necessary to obtain crystalline and stoichiometric $Ta_2O_5$ films even at substrate temperatures of 500°C.

## 2. Experimental:

All depositions and cathode-voltage-hysteresis experiments were performed in a Leybold Heraeus A400-VL laboratory magnetron sputtering device equipped with a 70 $dm^3$ chamber and a Leybold Turbovac 361 turbo pump (nominal pumping speed of 345 l/sec (nitrogen)). A tantalum target with a diameter of 75 mm (thickness 6mm) and a purity of 99.9% was used and powered by a Leybold/ELAN SSV1.8kW/2-27 DC generator in constant current mode at $I_m$ = 0.50, 0.75 and 1.00 A.

In order to decrease the radial magnetic field strength B at the target surface from 90 to 60 and 45 mT, austenitic steel plates were used as spacers between the cathode and the permanent magnets, see figure 1. The magnetic field strength was measured using a portable LakeShore 410 Gauss-meter.

Prior to the hysteresis experiments and without ignited plasma, we have calibrated the flows of Ar ($f_{Ar}$) and $O_2$ ($f_{O2}$) to reach a constant total pressure of 0.4 Pa. Therefore, $f_{O2}$ was stepwise increased from 0 to 23.7 sccm to linearly increase the $O_2$ partial pressure while simultaneously $f_{Ar}$ was decreased from 17 to 0 sccm. 30 separate data points were recorded per voltage hysteresis for increasing and subsequently decreasing flow rate Γ, with

$$\Gamma = \frac{f_{O2}}{f_{Ar} + f_{O2}} \times 100 .$$

The holding time for every single point was 5 minutes to approach equilibrium discharge conditions.

Tantalum oxide films were deposited on Si stripes ((100) orientation, 20x7x0.38 $mm^3$) using a target current of 0.75 A, a radial magnetic field strength of 60 mT, and flow rates of Γ = 50, 77 and 100%. The substrate temperature was set to 500 °C and the distance to the target surface was 4.5 cm.

Structural investigations were conducted by X-ray diffraction (XRD) analysis in Bragg Brentano geometry using a Brucker D8 diffractometer equipped with a CuKα radiation source. The chemical composition was determined by elastic recoil detection analysis (ERDA) using $Cl^{7+}$ ions with an acceleration voltage of 35 MeV by evaluating the resulting spectrum according to Barrada et al. [21]. Fracture cross-sectional scanning electron microscopy (SEM) investigations with an acceleration voltage of 15 kV were conducted with a Zeiss EVO50 for coating morphology studies and thickness evaluations. Prior to these SEM investigations, a thin Au-layer was deposited to increase the conductivity of our samples.

Energies of formation and density of states (DOS) of TaO (NaCl, Fm−3m (225), [22]), $TaO_2$ (rutile, P42/mnm (136), [23]) and $Ta_2O_5$ (orthorhombic structure proposed by Lehovec [24]) were calculated using the Vienna ab initio software package (VASP) [25,26] employing projector augmented wave pseudopotentials [27] and the generalized gradient approximation [28]. The unit cells of TaO, $TaO_2$ and $Ta_2O_5$ contained 8, 6, and 14 atoms, respectively and the used cut-off energy of 800 eV and more than 2000 k-points·atom ensure a minimum accuracy of 1 meV/atom. The energies of formation were calculated as,

$$E_f = E_{TaOx} - \frac{E_{Ta} + \frac{x}{2} \times E_{O2}}{1+x}$$

where $E_{TaOx}$, $E_{Ta}$ and $E_{O2}$ are the total energies of the crystalline $TaO_x$, bcc Ta and molecular $O_2$, respectively.

## 3. Results and Discussion:

### *3.1. Voltage hysteresis*

Figure 2 shows a voltage hysteresis, at a target current of $I_m$ = 0.75 A and a magnetic field strength of B = 90 mT as a function of the oxygen gas flow Γ. Between Γ = 0 and ~65% (first regime, (a)) a voltage peak of ~475 V can be observed at about Γ = 50%. The increase to this peak value can be attributed to the linear reduction of the Ar flow, oxygen gettering by evaporated metal atoms [29], chemisorption and implantation of oxygen atoms at the target [29,30], and also to the constriction of the discharge region due to partial poisoning [19]. At the peak, the discharge voltage and the ion density have reached the optimum conditions for restricting further oxidation of the target surface and the constriction of the discharge area has reached its maximum. If even more oxygen is introduced to the chamber, the target coverage increases, the discharge area broadens again and the voltage decreases until the poisoning processes are completed for the whole target surface. Subsequently stable discharge conditions, characterized by an almost constant voltage regime (b), can be observed up to $\Gamma = 90\%$, figure 2, and indicate that the formed compound is stable. For higher flow rates, between $\Gamma =$ 90 and 100% (c), the third regime is characterized again by an increase of the discharge voltage, due to the formation of a different compound layer at the target surface further influencing the target resistance. Upon decreasing the $O_2$ flow rate the hysteresis can be clearly identified for regime (a) and (c) in figure 2. Moreover, a non-stable DC glow discharge at $\Gamma = 100\%$ suggests a close to stoichiometric $Ta_2O_5$ compound formed on the target surface with a low conductivity, whereas the stable conditions within regime (b) suggest that the target surface is fully covered by a conductive substoichiometric $TaO_x$ phase.

Possible candidates for substoichiometric $TaO_x$ phases are metastable TaO and $TaO_2$ [31]. Therefore, ab initio calculations were performed to obtain information on the conductivity, stability and affinity to oxygen. These calculations yield energies of formation of −1.192 eV/atom (−230 kJ/mol) for TaO, −2.990 eV/atom (−865 kJ/mol) for $TaO_2$ and −3.158 eV/atom (−1079 kJ/mol) for $TaO_{2.5}$ (for comparison with the other phases we have normalized $Ta_2O_5$ to $TaO_{2.5}$). These results are in excellent agreement with experimentally obtained values of −1023 kJ/mol [32] and −1017 kJ/mol [33] for $TaO_{2.5}$. According to our calculations, the energy gain from TaO and $TaO_2$ towards the stable $TaO_{2.5}$ is 849 kJ/mol and 214 kJ/mol for TaO and $TaO_2$, respectively. The DOS for these three phases, figures. 3a, b, and c, indicate a shift in the (pseudo) bandgap (marked by the black arrow) towards the Fermi level with increasing oxidation state of Ta from TaO to $TaO_2$ to $Ta_2O_5$, respectively. For TaO and $TaO_2$ there are occupied states at the Fermi level suggesting conductivity. On the other hand, $Ta_2O_5$ exhibits a bandgap (~2.5 eV) at the Fermi-level, characteristic for an insulating material (please note that density functional theory (DFT) calculations tend to underestimate the bandgap width).

Due to these calculations and the much higher energy of formation for $TaO_2$ as compared with TaO, combined with the result that regime (b) of figure2 is characterized by a stable discharge over a wide range of oxygen flow rate, we propose that the properties of the tantalum oxide layer responsible for this regime are comparable to metastable $TaO_2$.

Moreover, the voltage increase (figure 2) from ~310V at $\Gamma$ = 0% to ~410V at $\Gamma$ = 75% to ~475V at $\Gamma$ = 100%, indicates decreasing secondary electron emission (SEE) yields for Ta, $TaO_x$ and $Ta_2O_5$ covered target surfaces, respectively, which is known for the Ta – $Ta_2O_5$ system [34].

*3.1.1. Current variation.* With increasing sputtering current $I_m$ from 0.50 to 0.75 to 1.00 A the three discharge regimes are clearly shifted to higher oxygen flow rates, figure 4. Furthermore, the regime with a stable discharge voltage (between the first and the second hysteresis) covers a broader oxygen flow rate with increasing sputtering current. For $I_m$ = 1 A, the second hysteresis effect due to the formation of an insulating $Ta_2O_5$ cannot be observed allowing for stable discharge conditions even for 100% of oxygen flow rate. Additionally, the second hysteresis, due to the formation of $Ta_2O_5$, clearly widens with decreasing target currents. This can be explained by the reduced particle bombardment (density) and energy caused by decreasing target current and voltage, and hence $Ta_2O_5$ stabilizes earlier during increasing flow rates, and (when formed) is stable to lower $\Gamma$ values during decreasing flow rates. The shift of the first hysteresis (formation of $TaO_x$) to lower oxygen flow rates with decreasing sputtering current is also based on the decreasing particle bombardment and consequently decreasing sputtering rate.

*3.1.2. Magnetic field strength variation.* The effect of progressing target erosion is investigated by modifying the magnetic field strength B at the surface (see experimental). Figures 5a and b show the discharge voltage behaviour for radial magnetic field strengths B of 45, 60, and 90 mT at a constant target current $I_m$ of 0.50 and 0.75 A, respectively. Increasing magnetic field strengths at constant target currents shifts the target voltage characteristics to lower values as more electrons are captured and more particles are ionized. Whereas the magnetic field strength variation (within the investigated ranges) has no significant influence on the position of the first hysteresis voltage peak, the second hysteresis is strongly influenced in shape and position. For both target current conditions of $I_m$ = 0.50 and 0.75 A no second hysteresis can be observed for the lower magnetic field strengths B of 45 and 60 mT, see figures 5a and b. This indicates that under these conditions no $Ta_2O_5$ compound layer is formed at the target surface. In this case we assume the cohesive energy of the present phase on the target surface, which is in general much lower for a metal than for the particular oxide, to play a major role. Consequently, mainly the density of the impinging ions is influencing the flow rate position of the Ta to $TaO_x$ hysteresis. Furthermore, the data also suggests that the second transition zone from $TaO_x$ to $Ta_2O_5$ is additionally to the density also influenced by the energy of the target impinging particles.

In agreement to the voltage dependence of the $TaO_x$ formation, which suggests decreasing SEE yields from Ta to $TaO_x$ to $Ta_2O_5$ (last paragraph in section 3.1) the above discussed magnetic field strength variation suggests decreasing secondary electron energies from Ta to $TaO_x$. Thus, the dependence of the discharge voltage on the magnetic field strength is more pronounced for the metallic Ta target surface at $\Gamma$ = 0% as compared with the oxidized state.

These investigations clearly demonstrate that the discharge and consequently the poisoning behaviour of the target are substantially influenced by the target erosion induced increase of the magnetic field strength, especially when operating at higher oxygen flow rates and lower sputtering currents.

*3.2. Film deposition*

Based on the previous studies we have chosen the settings of $I_m$ = 0.75 A and B = 60 mT for the preparation of tantalum oxide films using oxygen flow rates of 50, 77 and 100%, as these allow for a stable deposition process even at higher oxygen flow rates. The comparison of the cathode voltage and deposition rate as a function of the oxygen flow rate, figure 6, clearly exhibits a maximum deposition rate of 110 nm/min, when using the oxygen flow rate slightly above the peak voltage. Increasing the oxygen flow rate $\Gamma$ to 77 and 100% causes the deposition rate to decrease to 15 and 10 nm/min, respectively. The decrease from 110 to 15 nm/min can mainly be related to the different target surface condition, which is still metal dominated at the voltage peak for Γ = 50%, but completely covered by a $TaO_x$ compound layer for Γ = 77%. Further decreased deposition rates of 10 nm/min when using Γ = 100% are probably caused by different sputtering yields of Ar and $O_2$, as proposed by Yamamura et. al. [35], as well as due to changes in crystallinity and stoichiometry of the coatings, see next paragraph.

The coating prepared at Γ = 50% exhibits a dense columnar crystalline structure for the ~10 µm thick top region, preceded by a featureless, amorphous-like morphology up to a layer thickness of the first ~5 µm, figure 7a. This was confirmed when decreasing the deposition time from 120 to 5 min, as thereby the ~1 µm thin film exhibits only a featureless morphology, figure 7b. This featureless near-substrate region of the coating decreases with increasing oxygen flow rate and decreasing deposition rate, and is ~0.5 µm for Γ = 77% and essentially zero for Γ = 100%, see figures. 8a and b. XRD investigations, figure 9, confirm the crystalline nature of a highly texturized (110)/(200) orthorhombic $Ta_2O_5$ structure, in the outer most part of the coatings prepared at Γ = 50%, pattern (a). When decreasing the layer thickness to ~1 µm, only an X-ray amorphous response is obtained, pattern (b), in agreement to the SEM fracture cross section, figure 7b. The coatings prepared at higher oxygen flow rates, Γ = 77 and 100%, exhibit an orthorhombic $Ta_2O_5$ structure as well. However, with increasing oxygen flow rate the preferred orientation changes towards (001), see figure 9 pattern (c) and (d). The chemical compositions of our coatings, table 1, suggest that only for Γ = 100% a stoichiometric $Ta_2O_5$ with O/Ta = 2.5, is obtained, indicating that high oxygen partial pressures are needed to form crystalline stoichiometric $Ta_2O_5$ as proposed by Ritter [36]. The coatings synthesized at Γ = 50 and 77% are slightly substoichiometric, with an O/Ta ratio of 2.33 and 2.35, respectively.

## 4. Summary and conclusions

Voltage hysteresis of a Ta target in an $Ar/O_2$ atmosphere at three different sputtering currents and magnetic field strengths were investigated with respect to its target poisoning behaviour as well as resulting film structure and morphology. We indicated three main sputtering regions which are dominated by the target surface transition from Ta to $TaO_x$, a stable conductive $TaO_x$ region and finally the transition of $TaO_x$ to insolating $Ta_2O_5$. All three regions are strongly influenced by the cathode current used. Variations of the magnetic field strength mainly affect the second and third regime, whereas the first one does not change significantly. This can be related to the difference in cohesive energies of metallic Ta and Ta-oxide. XRD investigations of deposited Ta-oxides at 50, 77, and 100% oxygen flow rate exhibit a pentoxide structure. However, stoichiometric films could only be obtained at $\Gamma$ = 100%, whereas the O-to-Ta rate of the coatings prepared with $\Gamma$ = 50 and 77% is slightly sub-stoichiometric with 2.33 and 2.35 respectively. These coatings also exhibit an amorphous phase content next to the substrate interface.

Our investigations demonstrate the pronounced influence of the magnetic field strength, which may change due to e.g. target erosion, on the target poisoning and discharge behaviour, especially at high reactive gas flow ratios, which are necessary to obtain stoichiometric crystalline $Ta_2O_5$ coatings.

## 5. Acknowledgement

This work has been supported by the European Community as an Integrating Activity 'Support of Public and Industrial Research Using Ion Beam Technology (SPIRIT)' under EC contract no. 227012 as well as by the Start Program (Y 371) of the Austrian Science Fund (FWF) and the “Christian Doppler” research association.

**Figures:**

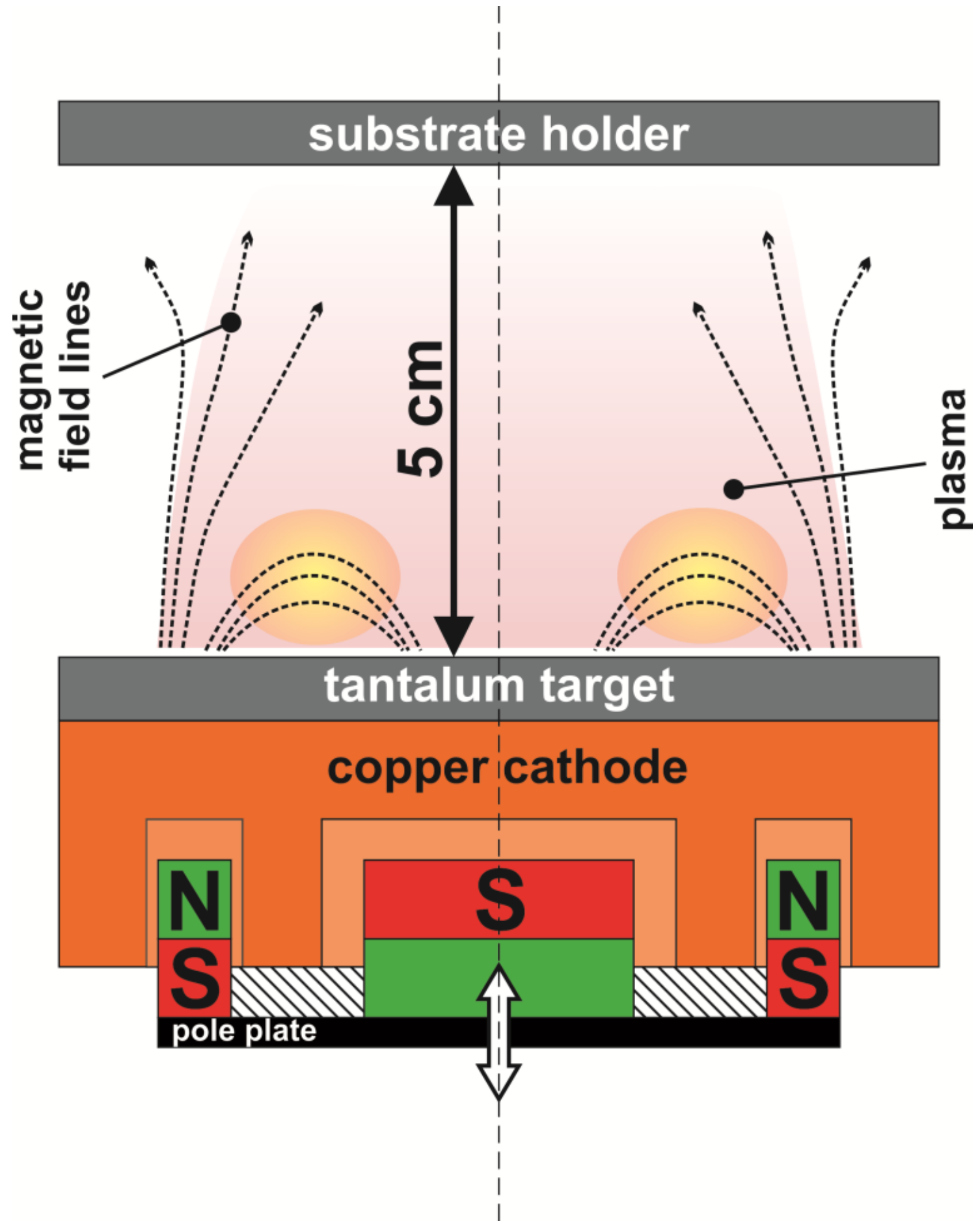

Figure 1: Sectional schematic of the used unbalanced magnetron system. The cross-striped area indicates the austenitic steel spacer sheets.

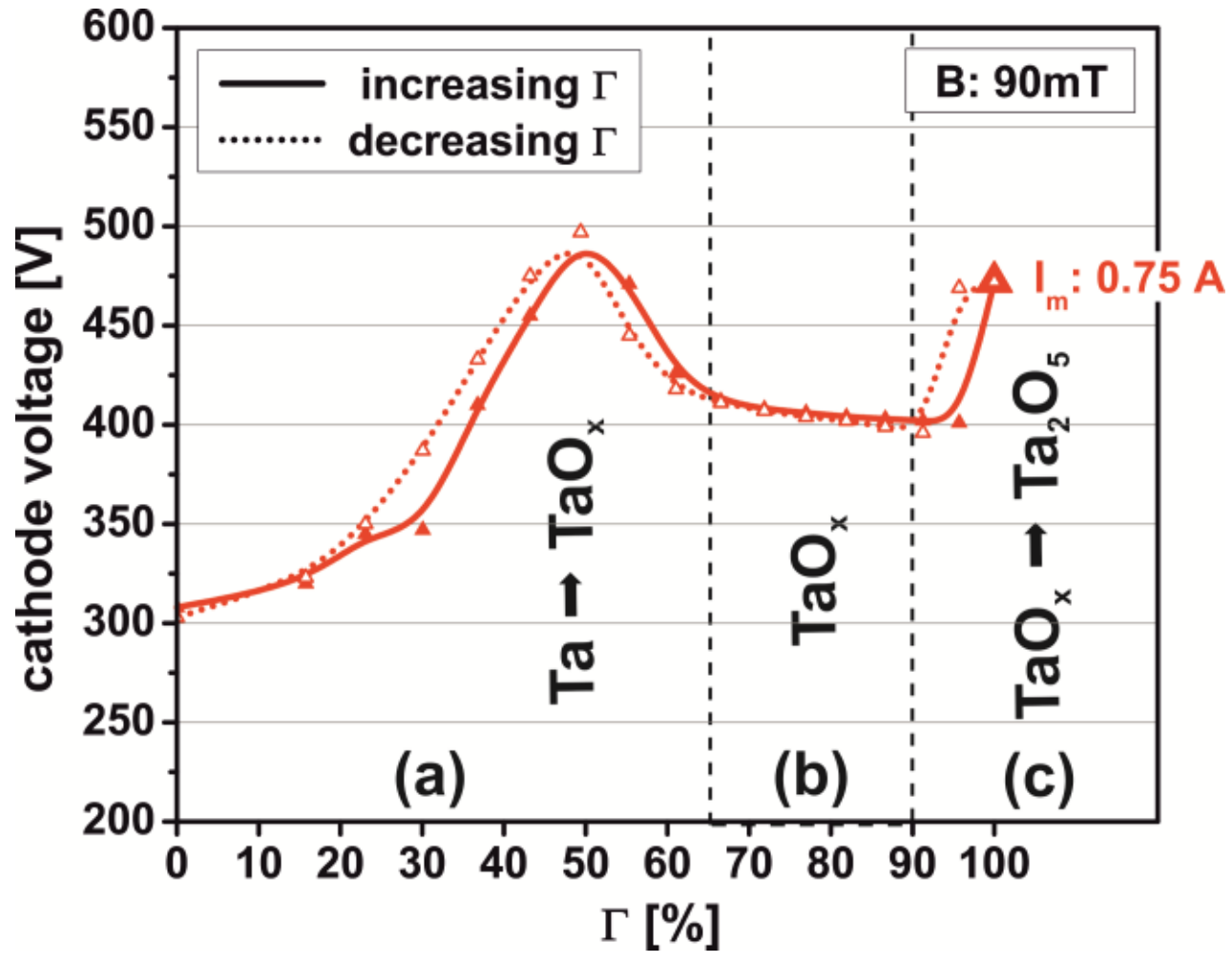

Figure 2: Cathode voltage hysteresis indicating the three main target poisoning regimes: (a) transition from Ta to $TaO_x$, (b) region of stable $TaO_x$, and (c) transition from $TaO_x$ to $Ta_2O_5$.

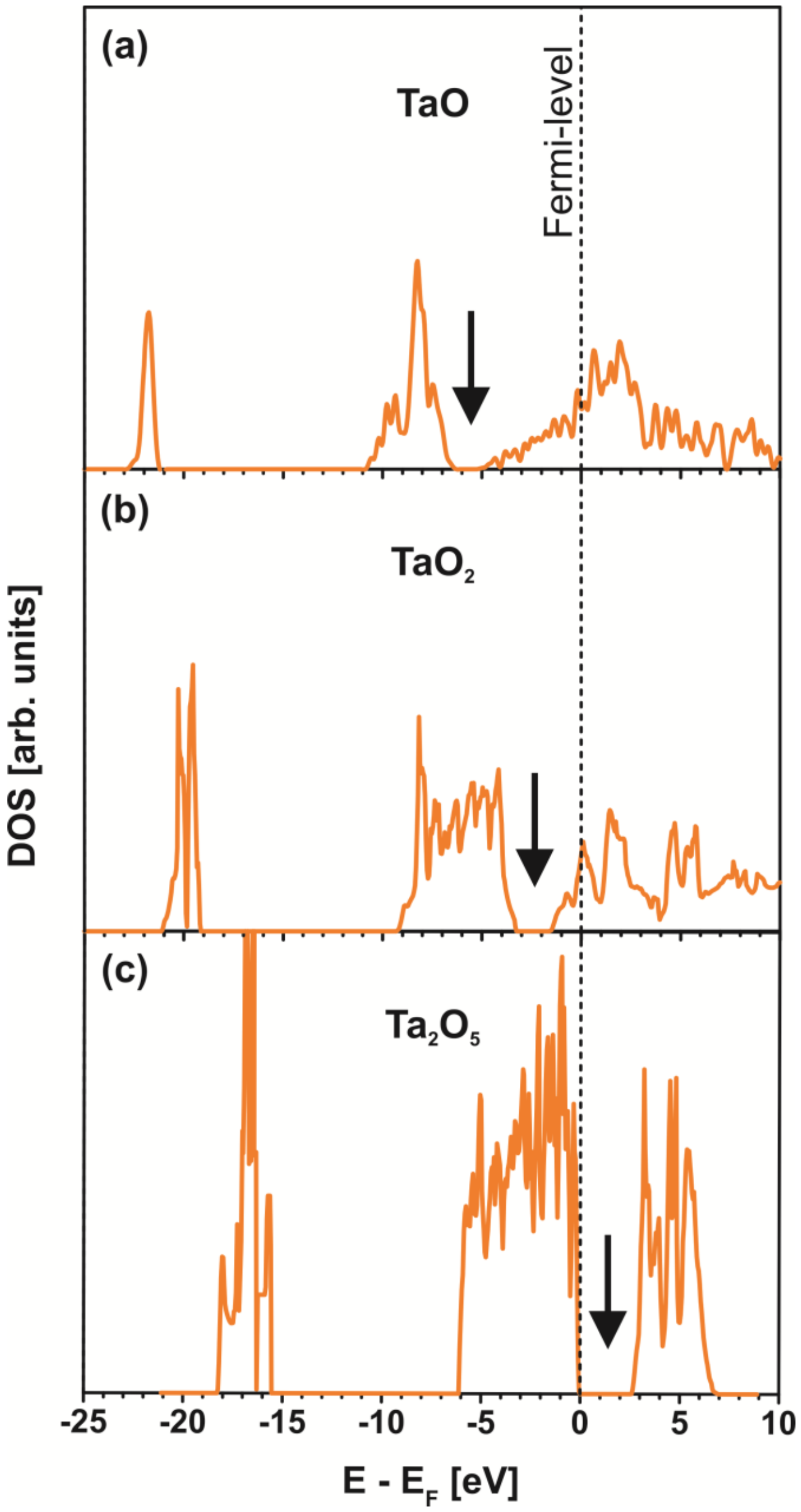

Figure 3: Total density of states (DOS) of (a) TaO, (b) $TaO_2$, and (c) $Ta_2O_5$. The vertical arrows indicate the (pseudo-) bandgap.

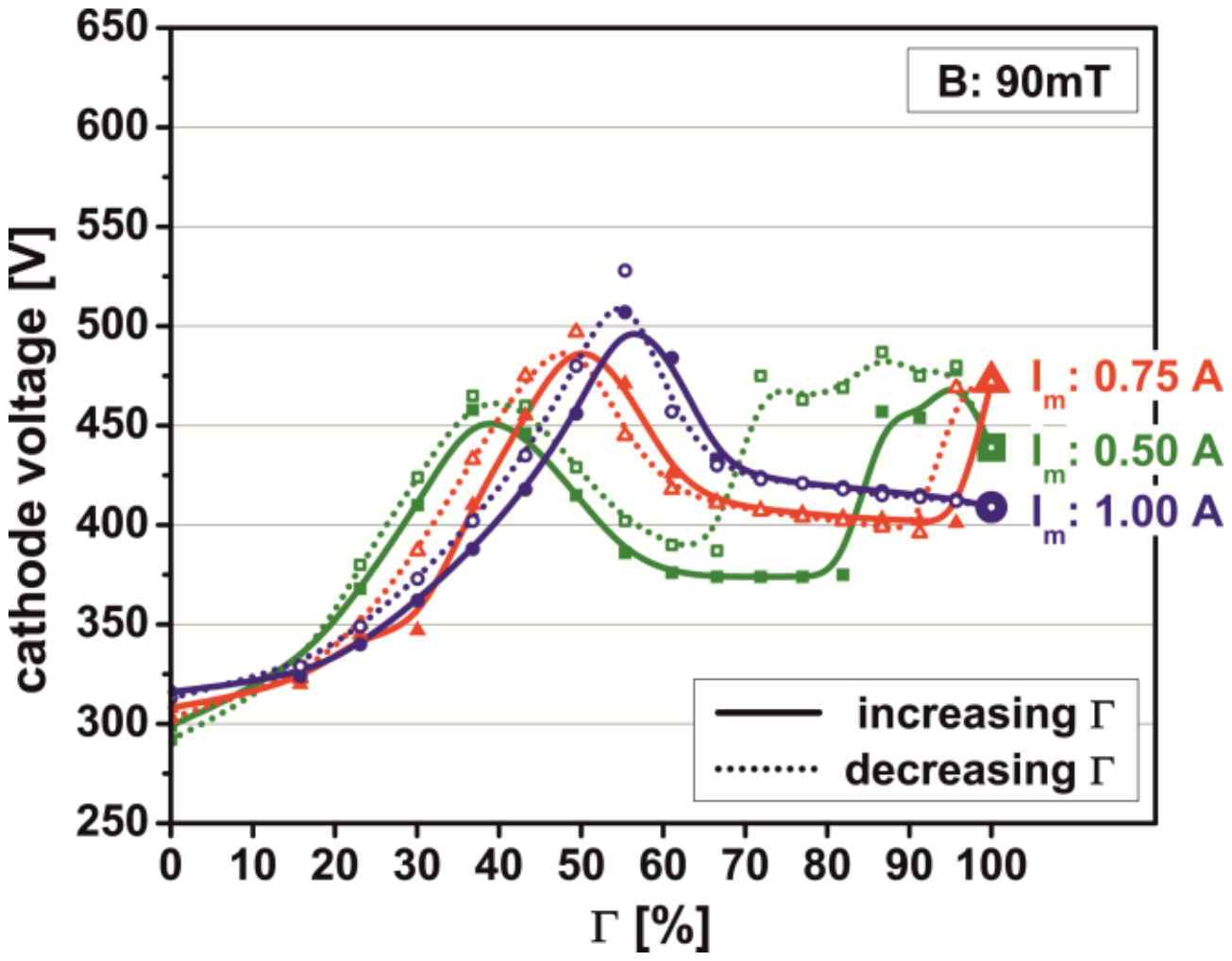

Figure 4: Cathode voltage hysteresis at a constant radial magnetic field strength of 90 mT and target currents of 0.50, 0.75, and 1.00 A.

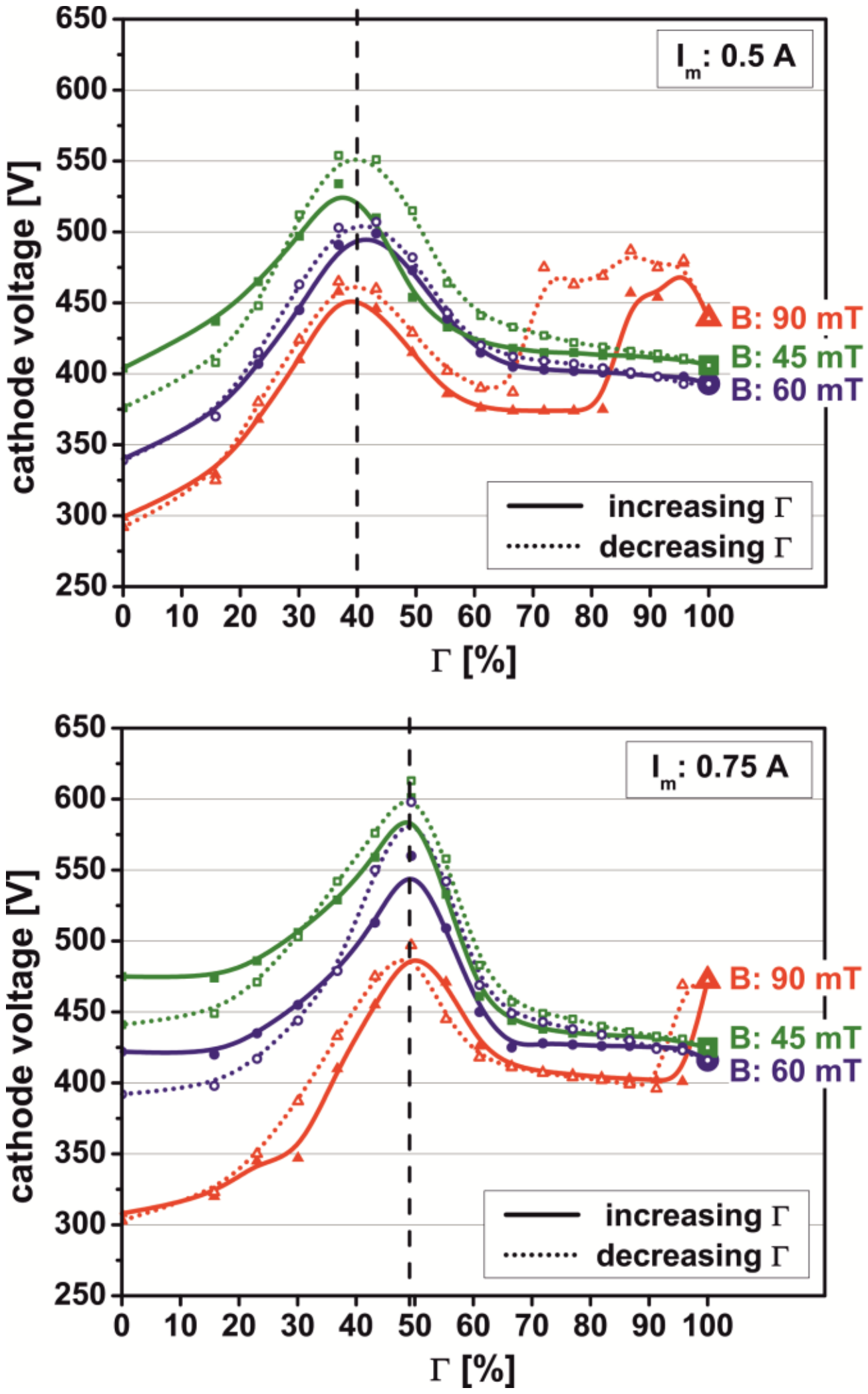

Figure 5: Cathode voltage hysteresis for magnetic field strengths of 45, 60 and 90 mT at target currents $I_m$ of (a) 0.50 and (b) 0.75 A.

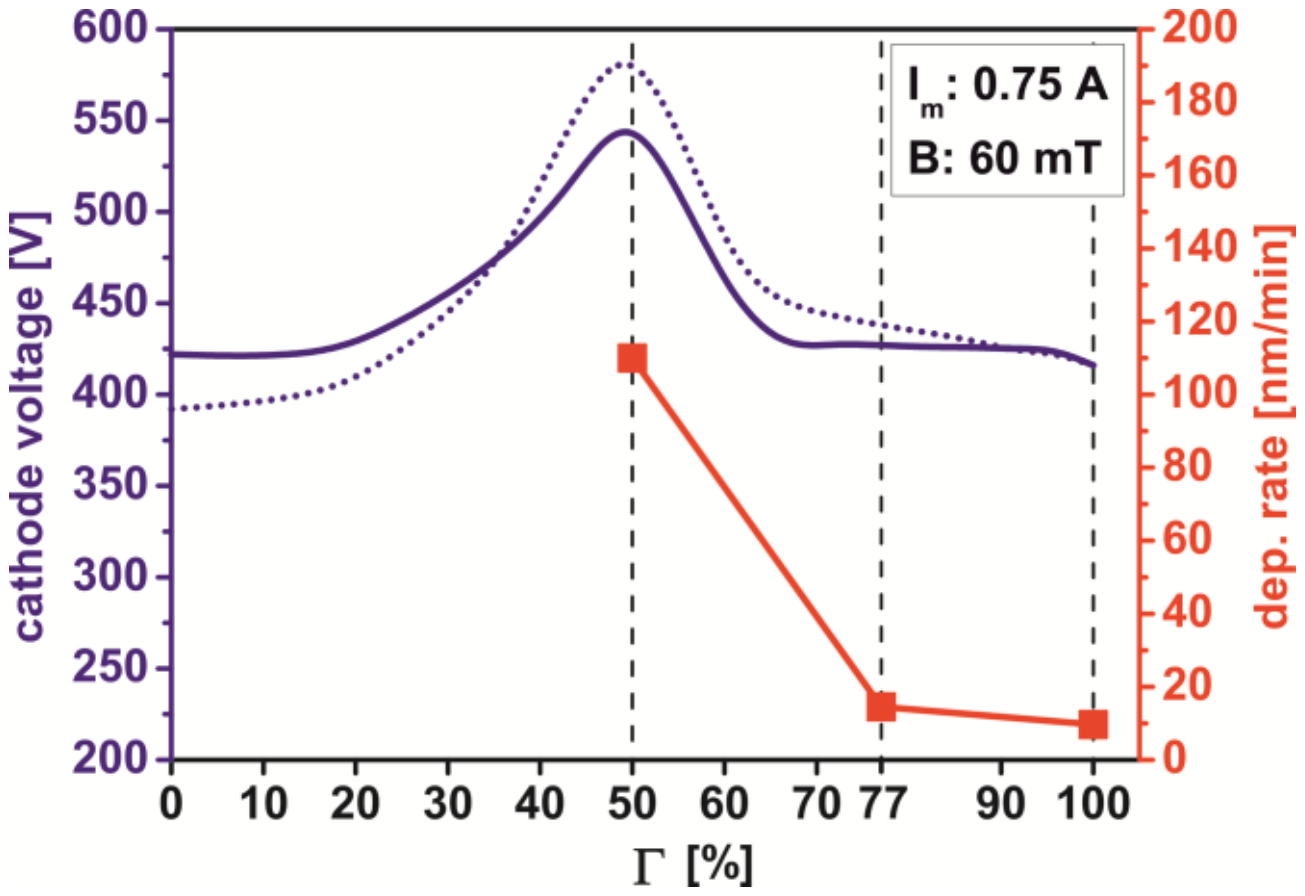

Figure 6: Cathode voltage hysteresis used for film deposition and resulting deposition rates for $I_m$ = 0.75 A and B = 60 mT.

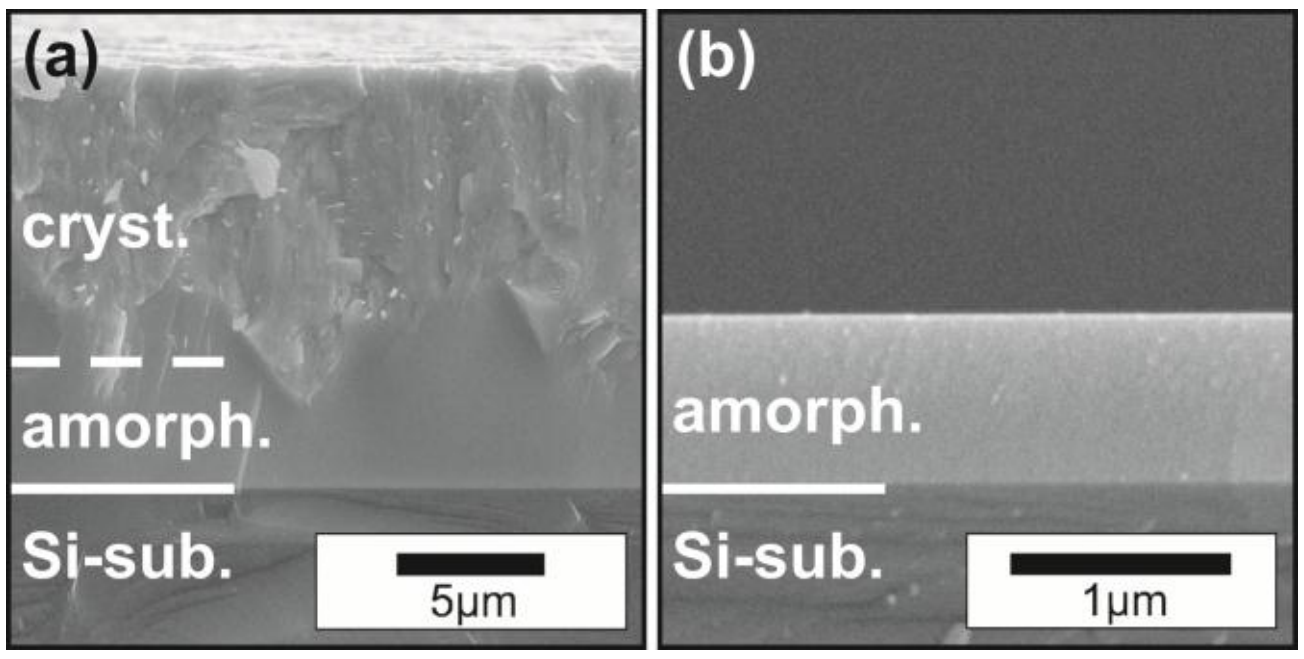

Figure 7: Cross-sectional SEM micrographs of tantalum oxide films deposited at Γ = 50% for deposition times of (a) 120 min and (b) 5 min.

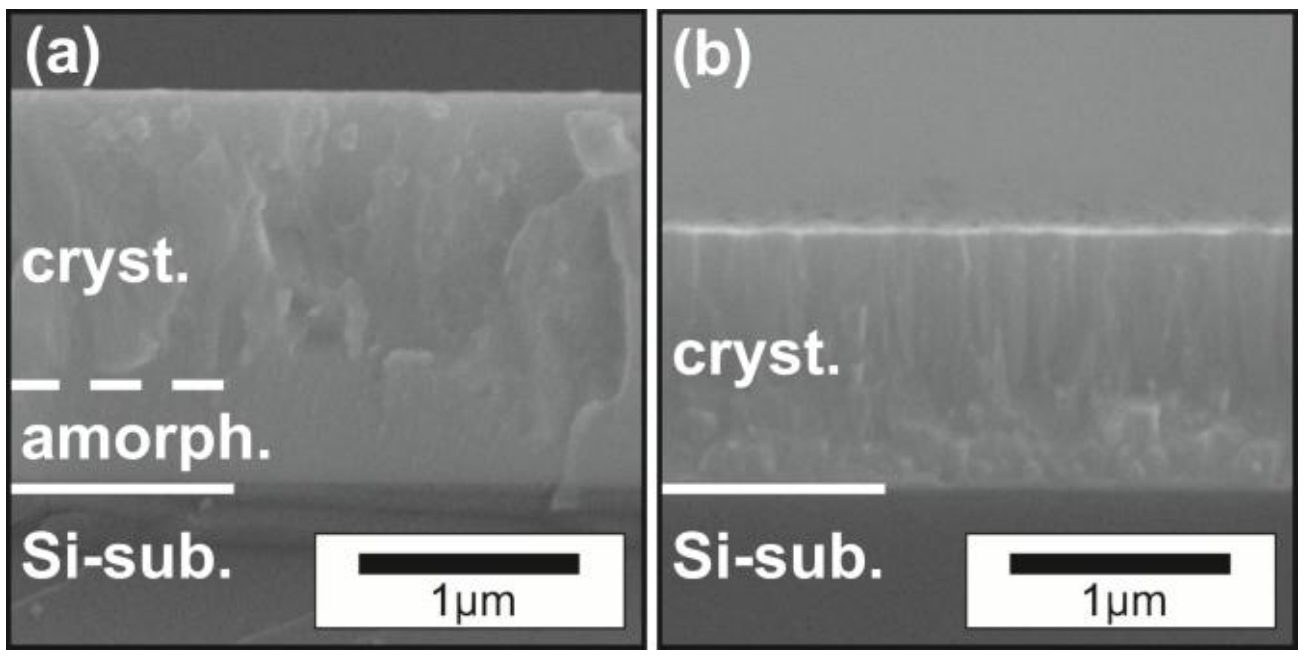

Figure 8: Cross-sectional SEM micrographs of tantalum oxide films deposited at (a) Γ = 77% and (b) 100% for a deposition time of 120 min.

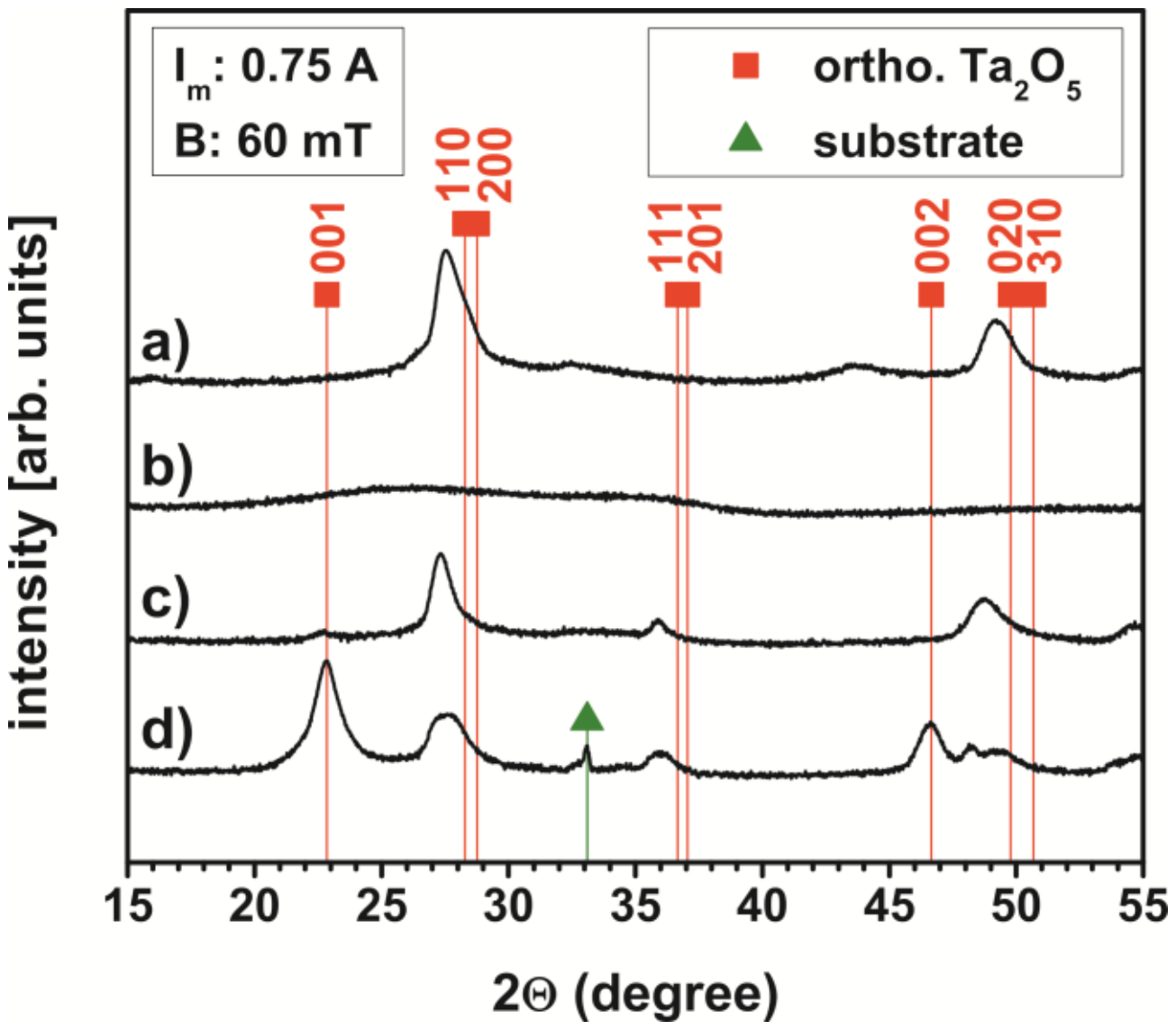

Figure 9: XRD patterns for tantalum oxide films deposited at (a) Γ = 50% for 120 min, (b) Γ = 50% for 5 min, (c) Γ = 77% for 120 min, and (d) Γ = 100% for 120 min (indexed after Lehovec et. al. [37]). The corresponding SEM cross-sections are shown in Figures. 7a, b and Figures. 8a, b, respectively.

## Tables

**Table 1.** ERDA evaluated chemical composition of $Ta_2O_5$ thin films

| Γ (%) | Ta (at%) | O (at%) | O/Ta (ratio) |
|---|---|---|---|
| 50 | 30.00 ± 0.1 | 70.00 ± 0.7 | 2.33 ± 0.03 |
| 77 | 29.83 ± 0.1 | 70.17 ± 0.7 | 2.35 ± 0.03 |
| 100 | 28.30 ± 0.1 | 71.70 ± 0.7 | 2.53 ± 0.03 |
| stoichiometric | 28.57 | 71.43 | 2.5 |